\documentclass[aps,prb,reprint,groupedaddress,superscriptaddress,amsmath,showpacs]{revtex4-1}
\usepackage{multirow}
\usepackage{tikz}
\usetikzlibrary{matrix,arrows,arrows.meta}

\begin{document}

\title{Raman phonons in the ferroelectric-like metal $\text{LiOsO}_\text{3}$}
\author{Feng\,Jin}
\affiliation{Department of Physics, Beijing Key Laboratory of Opto-Electronic Functional Materials $\&$ Micro-nano Devices, Renmin University of China, Beijing 100872, People's Republic of China}
\author{Anmin\,Zhang}
\affiliation{Department of Physics, Beijing Key Laboratory of Opto-Electronic Functional Materials $\&$ Micro-nano Devices, Renmin University of China, Beijing 100872, People's Republic of China}
\author{Jianting\,Ji}
\affiliation{Department of Physics, Beijing Key Laboratory of Opto-Electronic Functional Materials $\&$ Micro-nano Devices, Renmin University of China, Beijing 100872, People's Republic of China}
\author{Kai\,Liu}
\affiliation{Department of Physics, Beijing Key Laboratory of Opto-Electronic Functional Materials $\&$ Micro-nano Devices, Renmin University of China, Beijing 100872, People's Republic of China}
\author{Le\,Wang}
\affiliation{Beijing National Laboratory for Condensed Matter Physics, Institute of Physics, Chinese Academy of Sciences, Beijing 100190, People's Republic of China}
\author{Youguo\,Shi}
\affiliation{Beijing National Laboratory for Condensed Matter Physics, Institute of Physics, Chinese Academy of Sciences, Beijing 100190, People's Republic of China}
\author{Yong\,Tian}
\affiliation{Department of Physics, Beijing Key Laboratory of Opto-Electronic Functional Materials $\&$ Micro-nano Devices, Renmin University of China, Beijing 100872, People's Republic of China}
\author{Xiaoli\,Ma}
\affiliation{Department of Physics, Beijing Key Laboratory of Opto-Electronic Functional Materials $\&$ Micro-nano Devices, Renmin University of China, Beijing 100872, People's Republic of China}
\author{Qingming\,Zhang}
\email[Corresponding author: ]{qmzhang@ruc.edu.cn}
\affiliation{Department of Physics, Beijing Key Laboratory of Opto-Electronic Functional Materials $\&$ Micro-nano Devices, Renmin University of China, Beijing 100872, People's Republic of China}

\begin{abstract}
The novel ferroelectric-like structural transition observed in metallic $\text{LiOsO}_\text{3}$ [Y. Shi et al., Nat. Mater. 12, 1024 (2013)], has invoked many theoretical and experimental interests. In this work, we have performed polarized and temperature-dependent Raman scattering measurements on high-quality single crystal $\text{LiOsO}_\text{3}$ and identified Raman-active modes in both centrosymmetric phase (300 K, $R\bar{3}c$) and non-centrosymmetric phase (10 K, $R3c$). Only four phonon peaks are observed in the former phase, while there are twelve peaks in the latter phase because of the reduction of crystal symmetry. With the help of careful symmetry analysis and first-principles calculations, we can make a systematic assignment for the observed Raman modes in both phases. The significant changes in line-width and the continuous evolution of Raman frequencies with temperatures were observed for the $E_g$ modes around the transition temperature, which suggests that the ferroelectric-like structural transition is a continuous order-disorder transition. The result sheds light on the coexistence of ferroelectricity and metallicity in the compound.
\end{abstract}


\maketitle

\section{introduction}
Ferroelectricity has been widely studied since the first ferroelectric Rochelle-salt was discovered. Cochran\cite{Cochran1960} and Anderson\cite{Anderson1958} pointed out that ferroelectric instability originates from the approximate cancellation between short-range restoring force and dipole-dipole interaction, where the two terms favor high-temperature (HT) paraelectric and low-temperature (LT) ferroelectric phases, respectively. It was long believed that ferroelectricity in a metal is impossible because of the strong electronic screening for dipole-diploe interaction\cite{Lines2001}. However, in 1965 Anderson and Blount proposed the possible coexistence of ferroelectricity and metallicity in the framework of Landau theory\cite{Anderson1965}. They argued that the transition from paraelectric to ferroelectric phases in a metal can not be caused by strain. The transition must be continuous and accompanied by the emergence of a unique polar axis and disappearance of space inversion center. The search for ferroelectric metals has been ongoing for decades. Many compounds, such as A-15 superconductors, $\text{Cd}_\text{2}\text{Re}_\text{2}\text{O}_\text{7}$ and $\text{BaTiO}_{3-\delta}$, have been considered as potential ferroelectric metals\cite{Anderson1965,Sergienko2004,Kolodiazhnyi2010}. But further experiments\cite{Testardi1967,Ishibashi2010,Jeong2011} have shown that the transitions in the compounds do not really meet the criteria proposed by Anderson and Blount.
\par
In 2013, Y. Shi et al. reported\cite{Shi2013} the first ferroelectric-like metal $\text{LiOsO}_\text{3}$, in which a continuous structural phase transition from centrosymmetric $R\bar{3}c$ group to non-centrosymmetric $R3c$ group, occurs at 140 K. Neutron and X-ray diffraction experiments demonstrate that the transition involves the continuous changes of average positions of Li atoms along the c axis, accompanying with a subtle displacement of oxygen atom. This kind of transition is usually seen in \emph{insulating} compounds like $\text{LiNbO}_\text{3}$ and $\text{LiTaO}_\text{3}$, etc., and consequently produces ferroelectricity. Surprisingly, resistivity measurements confirm that both phases in $\text{LiOsO}_\text{3}$ are \emph{metallic}. The lacking of inversion center in the LT phase suggests the possibility of electric dipole. This points to the coexistence of ferroelectricity and metallicity in the LT phase. This ferroelectric-like transition seems to contradict with the soft mode mechanism proposed by Cochran\cite{Cochran1960} and Anderson\cite{Anderson1958}, since no effective dipole-dipole interaction can be expected in a metal. This may require a new mechanism for the ferroelectric transition in metals. And the related topics have attracted many theoretical interests\cite{Sim2014,Xiang2014,Liu2015,Giovannetti2014}. The theories of soft mode\cite{Sim2014}, order-disorder\cite{Liu2015} and the hybridization between $E_g$ orbits of Os and $p$ orbits of oxygen\cite{Giovannetti2014}, were proposed to understand the ferroelectric-like transition in $\text{LiOsO}_\text{3}$. Additionally, the weak coupling between electrons and softening phonons was also considered to be the possible driving force\cite{Puggioni2014}. So far, most of studies come from the theoretical side and experimental studies are highly required.
\par
In this paper, we have carried out polarized Raman measurements on high-quality single crystal $\text{LiOsO}_\text{3}$ and collected Raman spectra in the high-temperature (HT) centrosymmetric phase and low-temperature(LT) non-centrosymmetric phase. The observed Raman modes are carefully assigned with the aid of careful symmetry analysis and first-principles calculations. Only four modes are observed in the HT phase, while there are more than 12 modes in the LT phase due to the symmetry reduction. The modes are related to the vibrations of Li atom and the rotating, bending and anti-symmetric stretching of $\text{OsO}_\text{6}$ octahedra. Around the transition temperature, several modes show significant increases in line-width and continuous changes in Raman shift, which may imply a continuous order-disorder mechanism for the transition.

\begin{table}[t]
\caption{\label{table1} Wyckoff positions, atomic site symmetries for
         $\mathrm{LiOsO_3}$ at HT phase (space group $R\bar{3}c$, No.167) 
         and LT phase (space group $R3c$, No.161).}
\begin{ruledtabular}
\begin{tabular}{cccc}
       \textrm{~}&Atom&Wyckoff position&Site symmetry\\ \hline
       space group $R\bar{3}c$ & Li & $6a$  & $D_3$\\
       (No.167)                & Os & $6b$  & $S_6$\\
       ~                       & O  & $18e$ & $C_2$\\ \hline
       space group $R3c$       & Li & $6a$  & $C_3$\\
       (No.161)                & Os & $6a$  & $C_3$\\
       ~                       & O  & $18b$ & $C_1$\\

\end{tabular}
\end{ruledtabular}
\end{table}

\section{Experiments and methods}
The crystals used in our measurements were grown under high pressure and carefully characterized by X-ray diffraction and neutron diffraction\cite{Shi2013}. The cleaved samples were placed in a UHV cryostat with a vacuum of $\sim$10$^{-8}$ mbar. Raman spectra were collected with a LABRAM HR800 system, which is equipped with a single grating of 800 mm focus length and liquid-nitrogen-cooled CCD. About 1 mW of laser power at 632.8 nm was focused into a spot with a diameter of $\sim$~5 microns on the sample surface. The principal axes of the crystals are hard to be identified by conventional structural methods, since the typical dimension of available crystal surfaces is only tens of microns after cleavage. The temperature increase by laser heating has been monitored and calibrated by Stokes-antiStokes relationship and the transition temperature determined by thermodynamics measurements.
\par
	To determine the vibrational properties of $\text{LiOsO}_\text{3}$, first-principles calculations were implemented with the Vienna Ab initio Simulation Package (VASP)\cite{Kresse1993,*Kresse1996a,*Kresse1996}. The projector augmented wave method\cite{Bloechl1994,*Kresse1999} was used to describe the core electrons. For the exchange-correlation potential, the generalized gradient approximation (GGA) of Perdew-Burke-Ernzerh\cite{Perdew1996} was adopted. The kinetic energy cutoff of the plane-wave basis was set to be 600 eV. The simulations were carried out with  a rhombic supercell  containing two Li atoms, two Os atoms, and six O atoms. The experimental lattice parameters at 300 K and 10 K\cite{Shi2013} were utilized for the starting structures of the high-temperature and low-temperature phases, respectively. For the Brillouin zone sampling, an 8$\times$8$\times$8 k-point mesh was employed. The Gaussian smearing with a width of 0.05 eV was used around the Fermi surface. In structure optimization, both cell parameters and internal atomic positions were allowed to relax until the forces were smaller than 0.005 eV/$\mathring{A}$. Once the equilibrium structure was obtained, the vibrational frequencies and polarization vectors at Brillouin zone center were calculated by using the dynamic matrix method.

\begin{table}[b]
\caption{\label{table2} Normal vibrations of $\mathrm{OsO_6}$ octahedra with different site symmetries, and the transformations between them. The correspondence between normal vibrations in the HT (space group $R\bar{3}c$, point group $D_{3d}$) and LT (space group $R3c$, point group $C_{3v}$) phases, is also given. The mode classification of $\mathrm{LiOsO_3}$ is summarized at the table bottom.}
\begin{ruledtabular}
\begin{tabular}{c|cccc}
  ~ & Free ion~~~ & Site symmetry            & Unite cell & Unite cell\\
  ~ & symmetry    & $D_3$ ($\mathrm{OsO_6}$) & symmetry   & symmetry\\
  ~ & $O_h$       & $S_6$(Li)                & $D_{3d}$   & $C_{3v}$\\\hline
  \multirow{8}{*}{$\mathrm{OsO_6}$}
\begin{picture}(0,0)(-18,120)
\begin{tikzpicture}
[scale=.8,auto=left]
\node (a) at (-4,1) {3\,$F_{1u}$};
\node (b) at ( -4,0) {1\,$F_{2u}$};
\node (c) at ( -1.5,1.5) {3\,$A_{2}$};
\node (d) at ( -1.5, 0.5) {4\,$E$};
\node (e) at ( -1.5, -0.5) {1\,$A_{1}$};
\node (f) at (-1.5, -1.5) {1\,$A_{u}$};
\node (g) at (-1.5, -2.5) {1\,$E_{u}$};
\node (h) at (1.1,2) {3\,$A_{2g}$};
\node (i) at (1.1,1) {4\,$A_{2u}$};
\node (j) at ( 1.1,0) {4\,$E_{g}$};
\node (k) at ( 1.1,-1) {6\,$E_{u}$};
\node (l) at ( 1.1, -2) {1\,$A_{1g}$};
\node (m) at ( 1.1, -3) {2\,$A_{1u}$};
\node (n) at (3.3, 1.2) {5\,$A_{2}$};
\node (o) at (3.3, -0.5) {5\,$A_{1}$};
\node (p) at (3.3,-2.2) {10\,$E$};
\draw [arrows={-Stealth[length=5pt,inset=2pt]}] (-3.4,1.1) -- (-2,1.4);
\draw [arrows={-Stealth[length=5pt,inset=2pt]}] (-3.4,0.9) -- (-2,0.6);
\draw [arrows={-Stealth[length=5pt,inset=2pt]}] (-3.4,0.1) -- (-2,0.4);
\draw [arrows={-Stealth[length=5pt,inset=2pt]}] (-3.4,-0.1) -- (-2,-0.40);

\draw [arrows={-Stealth[length=5pt,inset=2pt]}] (-1,1.6) -- (0.4,2);
\draw [arrows={-Stealth[length=5pt,inset=2pt]}] (-1,1.4) -- (0.4,1);
\draw [arrows={-Stealth[length=5pt,inset=2pt]}] (-1,0.6) -- (0.4,0.1);
\draw [arrows={-Stealth[length=5pt,inset=2pt]}] (-1,0.4) -- (0.4,-0.90);

\draw [arrows={-Stealth[length=5pt,inset=2pt]}] (-1,-0.5) -- (0.4,-1.9);
\draw [arrows={-Stealth[length=5pt,inset=2pt]}] (-1,-0.7) -- (0.4,-2.8);
\draw [arrows={-Stealth[length=5pt,inset=2pt]}] (-1,-1.4) -- (0.4,0.9);
\draw [arrows={-Stealth[length=5pt,inset=2pt]}] (-1,-1.6) -- (0.4,-3.0);
\draw [arrows={-Stealth[length=5pt,inset=2pt]}] (-1,-2.5) -- (0.4,-1);

\draw [arrows={-Stealth[length=5pt,inset=2pt]}] (1.6,2) -- (2.7,1.3);
\draw [arrows={-Stealth[length=5pt,inset=2pt]}] (1.6,0.9) -- (2.7,-0.4);
\draw [arrows={-Stealth[length=5pt,inset=2pt]}] (1.6,0) -- (2.7,-2);
\draw [arrows={-Stealth[length=5pt,inset=2pt]}] (1.6,-1) -- (2.7,-2.2);
\draw [arrows={-Stealth[length=5pt,inset=2pt]}] (1.6,-2) -- (2.7,-0.6);
\draw [arrows={-Stealth[length=5pt,inset=2pt]}] (1.6,-3) -- (2.7,1.1);
\end{tikzpicture}
\end{picture}
\\\\\\\\\\\\
\cline{1-2}\\
\multirow{2}{*}{$\mathrm{Li}$}
\\\\\\
\hline
\multicolumn{5}{c}{Modes classification}\\
\multicolumn{5}{l}{$R\bar{3}c$ :~$\Gamma_R$=\,$A_{1g}$+4$E_{g}$~,~$\Gamma_{IR}$=3$A_{2u}$+5$E_{u}$,}\\
\multicolumn{5}{l}{~~~~~~~~$\Gamma_{Acoustic}$=$A_{2u}$+$E_{u}$,~$\Gamma_{S}$=3$A_{2g}$+2$A_{1u}$.}\\
\multicolumn{5}{l}{$R3c$ :~$\Gamma_{R/IR}$=4$A_{1}$+9$E$~,~$\Gamma_{Acoustic}$=$A_{1}$+$E$~,~$\Gamma_{S}$=5$A_{2}$.}\\
\end{tabular}
\end{ruledtabular}
\end{table}

\begin{figure}[b] \scalebox{0.43}{\includegraphics*{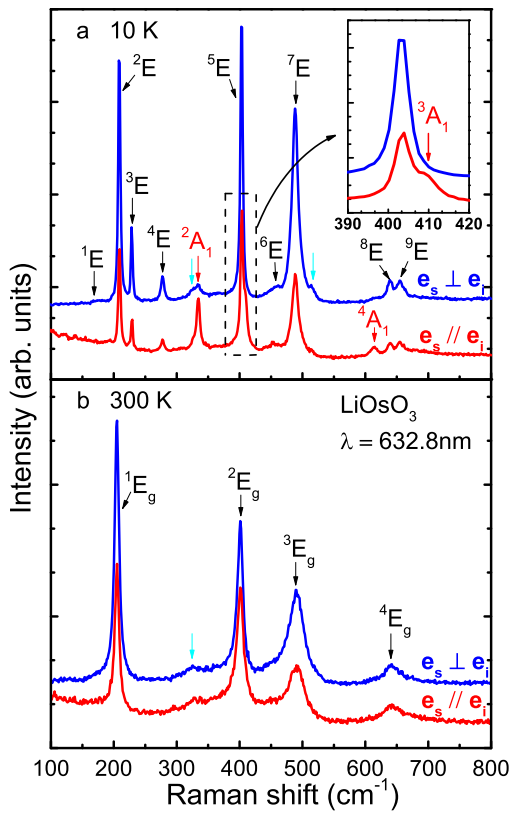}}
\caption{\label{fig1} Raman spectra of $\text{LiOsO}_\text{3}$ crystals at 10 K(a) and 300 K(b). Labels $e_i$ and $e_s$ denote the polarizations of the incident and scattered light. $A$ and $E$ modes are marked by black and red arrows, respectively, and the cyan arrows indicate the weak modes which may originate from impurities.}
\end{figure}

\section{Results and discussions}
Similiar to $\text{BiFeO}_\text{3}$\cite{Hermet2007}, without any tilting of $\text{OsO}_\text{6}$ octahedra, $\text{LiOsO}_\text{3}$ has a $Pm\bar{3}m$ cubic structure. If there appears any tilting of $\text{OsO}_\text{6}$ octahedra around [111] direction of the pseudo-cubic, the $R\bar{3}c$ structure will be stabilized\cite{Shi2013,Glazer1972}. In this case the tilt angle can be estimated from experimental cell parameters\cite{Moreau1970,Shi2013}, and is about 20.1$^\circ$ at 300 K. The angle is close to those in $\text{LiTaO}_\text{3}$ ($\sim$23$^\circ$) and $\text{LiNbO}_\text{3}$ ($\sim$23.5$^\circ$)\cite{Moreau1970},  but larger than the value of 14.11$^\circ$ given by first-principles calculations\cite{Sim2014}. The detailed information of the rhombohedral structure are summarized in Table~\ref{table1} and the main structural feature is dominated by $\text{OsO}_\text{6}$ octahedra. This allows to divide all the normal vibrations into two categories: vibrations of Li against  $\text{OsO}_\text{6}$ octahedra, and internal vibrations of $\text{OsO}_\text{6}$ octahedra. An isolated $\text{OsO}_\text{6}$ octahedron has a point-group symmetry of $O_h$, allowing six species of normal vibrations: $A_{1g}$, $E_{g}$, $F_{1u}$, $F_{1u}$, $F_{2g}$, $F_{2u}$\cite{Herzberg1945}. This set of normal vibrations are further reduced to 2$F_{1u}$+$F_{2u}$ because the atoms in equivalent positions in neighboring cells must perform the same vibrational motion\cite{Last1957}. If taking into account the $F_{1u}$ translational vibration, the total normal vibrations of  $\text{OsO}_\text{6}$  octahedra are 3$F_{1u}$+$F_{2u}$, which are further split into $A_{1}$, $A_{2}$ and $E$ vibrations in the HT phase since the site symmetry of octahedra is reduced to $D_{3}$. Under the $D_{3d}$ symmetry of the HT phase, $A_{1}$, $A_{2}$ and $E$ vibrations are transformed to $A_{2g}$, $A_{2u}$, $E_{g}$, $E_{u}$, $A_{1g}$, and $A_{1u}$\cite{Rousseau1981}. The transformation paths are illustrated in Table~\ref{table2}. On the other hand, Li atoms in the HT phase are located at $S_{6}$ sites. Similar analysis gives the vibrations of $A_{1u}$+$A_{2u}$+2$E_{u}$. Thus, the total optical modes in the HT phase are $A_{1g}$+3$A_{2g}$+2$A_{1u}$+3$A_{2u}$+4$E_{g}$+5$E_{u}$, where $A_{1g}$ and $E_{g}$ modes are Raman-active, $A_{2u}$ and $E_{u}$ infrared-active, and $A_{2g}$ and $A_{1u}$ silent. For the $C_{3v}$ symmetry of LT phase,
through the correspondence between the modes of both phases\cite{Rousseau1981}, we have 4$A_{1}$+5$A_{2}$+9$E$ optical vibrational modes, where $A_{1}$ and $E$ modes are both Raman- and infrared-active, and $A_{2}$ modes are silent.

\begin{table}[b]
\caption{\label{table3} Experimental and calculated mode frequencies (in cm$^{-1}$),
                        optical activity of the optical phonon modes of $\mathrm{LiOsO_3}$.
                        The asterisk($\ast$) indicates low intensity and R, IR
                        and S denote Raman, Infrared and Silent, respectively.}
\begin{ruledtabular}
\begin{tabular}{cccc||cccc}
\multicolumn{4}{c||}{Low temperature phase} & \multicolumn{4}{c}{High temperature phase}\\\hline
 Mode & Cal. & Exp. & Active & Mode & Cal. & Exp. & Active\\
\hline
$^{1}\!{A_1}$ & 212 & -\,-  & R/IR & $^{1}\!{A_{2u}}$ & -138 &      & IR\\
$^{2}\!{A_1}$ & 328 & 334.1 & R/IR & $^{2}\!{A_{2u}}$ & 315  &      & IR\\
$^{3}\!{A_1}$ & 412 & 409.7 & R/IR & $^{1}\!{A_{1g}}$ & 415  & -\,- & R\\
$^{4}\!{A_1}$ & 607 & 613.9 & R/IR & $^{3}\!{A_{2u}}$ & 603  &      & IR\\
$^{1}\!{A_2}$ & 159 &       & S    & $^{1}\!{A_{2g}}$ & -87  &      & S\\
$^{2}\!{A_2}$ & 278 &       & S    & $^{1}\!{A_{1u}}$ & 223  &      & S\\
$^{3}\!{A_2}$ & 357 &       & S    & $^{2}\!{A_{1u}}$ & 363  &      & S\\
$^{4}\!{A_2}$ & 408 &       & S    & $^{2}\!{A_{2g}}$ & 392  &      & S\\
$^{5}\!{A_2}$ & 666 &       & S    & $^{3}\!{A_{2g}}$ & 666  &      & S\\
$^{1}\!{E}$   & 176 &174$^\ast$&R/IR&$^{1}\!{E_u}$    & 172  &      & IR\\
$^{2}\!{E}$   & 202 & 208.6 & R/IR & $^{1}\!{E_g}$    & 200  &205.9 & R\\
$^{3}\!{E}$   & 245 & 228.6 & R/IR & $^{2}\!{E_u}$    & 231  &      & IR\\
$^{4}\!{E}$   & 282 & 277.4 & R/IR & $^{3}\!{E_u}$    & 280  &      & IR\\
$^{5}\!{E}$   & 380 & 403.2 & R/IR & $^{2}\!{E_g}$    & 380  &401.9 & R\\
$^{6}\!{E}$   & 448 & 456.8 & R/IR & $^{4}\!{E_u}$    & 488  &      & IR\\
$^{7}\!{E}$   & 471 & 488.2 & R/IR & $^{3}\!{E_g}$    & 515  &491.7 & R\\
$^{8}\!{E}$   & 609 & 638.9 & R/IR & $^{4}\!{E_g}$    & 607  &642.9 & R\\
$^{9}\!{E}$   & 643 & 655.1 & R/IR & $^{5}\!{E_u}$    & 644  &      & IR\\
\end{tabular}
\end{ruledtabular}
\end{table}
\par
Experimentally four Raman modes are observed in the HT phase (300 K), locating at 205.9 cm$^{-1}$, 401.9 cm$^{-1}$, 491.7 cm$^{-1}$ and 642.9 cm$^{-1}$ respectively (Fig.~\ref{fig1}b). The relative intensities of the four modes remain almost unchanged in both the cross ($e_i\perp e_s$) and the parallel ($e_i\parallel
e_s$) configurations, where $e_i$ and $e_s$ are the polarizations of incident and scattered light, respectively. For the orthogonal coordinate system, the Raman tensors for the irreducible representations of the $D_{3d}$/$C_{3v}$ point group are
\begin{center}
  $A_{1g}/A(z)=\left(
     \begin{array}{ccc}
       a&0&0\\
       0&a&0\\
       0&0&b\\
     \end{array}
     \right),$
  $E_g/E(y)=\left(
     \begin{array}{ccc}
      c&0&0\\
      0&-c&d\\
      0&d&0\\
     \end{array}
     \right),$
\end{center}

\begin{center}
  $E_g/E(-x)=\left(
     \begin{array}{ccc}
      0&~~-c~~&-d\\
      -c&~~0~~&0\\
      -d&~~0~~&0\\
     \end{array}
     \right).$
\end{center}

The Raman tensors require the intensity ratio of $A_{1g}$ to $E_{g}$ modes will be much smaller in the cross configuration than that in the parallel case. This suggests that the observed four modes belong to $E_{g}$ symmetry. In addition, there appears a small peak around 326 cm$^{-1}$, which is also seen in the LT phase. We will discuss it later.	

\begin{figure*}[t]
\scalebox{0.83}{\includegraphics*{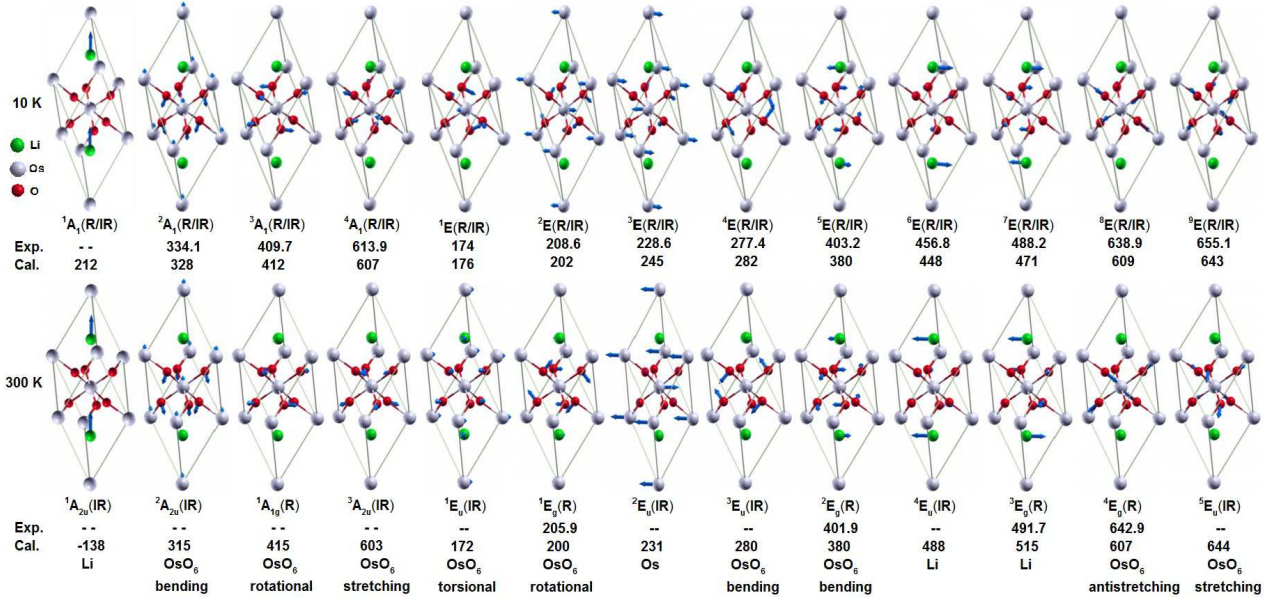}}
\caption{\label{fig2} Displacement patterns for all Raman-active modes of $\text{LiOsO}_\text{3}$. The mode symmetry is indicated below each vibrational pattern. The optical activity (IR=infrared active, R=Raman), the experimental and calculated mode energies (in cm$^{-1}$) and the main atomic displacements are also listed below. The atomic structures and vibrational displacement patterns were prepared with the XCRYSDEN program\cite{Kokalj2003}.}
\end{figure*}

\par
The above symmetry analysis gives thirteen Raman-active modes in the LT phase (4$A_{1}$+9$E$). Twelve peaks can be seen in the Raman spectra at 10 K (Fig.~\ref{fig1}a). Their positions are list in Table~\ref{table3}. Similarly, in the LT phase we can expect a much smaller intensity ratio of $A_{1}$  to $E$ modes in the cross channel. This allows us to attribute the modes at 334.1 cm$^{-1}$, 409.7 cm$^{-1}$ and 613.9 cm$^{-1}$ to $A_{1}$ symmetry. The assignment is well consistent with first-principles calculations (Table~\ref{table3}). It should be noted that the $A_{1}$  mode at 409.7 cm$^{-1}$ is close to the very strong $E$ mode at 403.2 cm$^{-1}$. This makes it hard to distinguish the $A_1$ mode from the $E$ mode in the cross channel. Fortunately we have successfully separated them in the parallel channel, as shown in the insets of Fig.~\ref{fig1}. The strongest $A_{1}$ mode at 334.1 cm$^{-1}$ has a residual intensity in the cross channel, which may come from intensity leakage due to the small mismatch between principal axes and the polarization of incident light. There is a small peak around 326 cm$^{-1}$, which also appears in the HT phase as mentioned above. This implies that the peak may originate from domain boundaries, defects or impurities. The peak at 516 cm$^{-1}$ may have the similar origin, since the calculations show no Raman or infrared-active modes in the frequency range. The rest nine peaks are identified as $E$ modes, in accord with symmetry analysis.

\begin{figure}[b]
\scalebox{0.45}{\includegraphics*{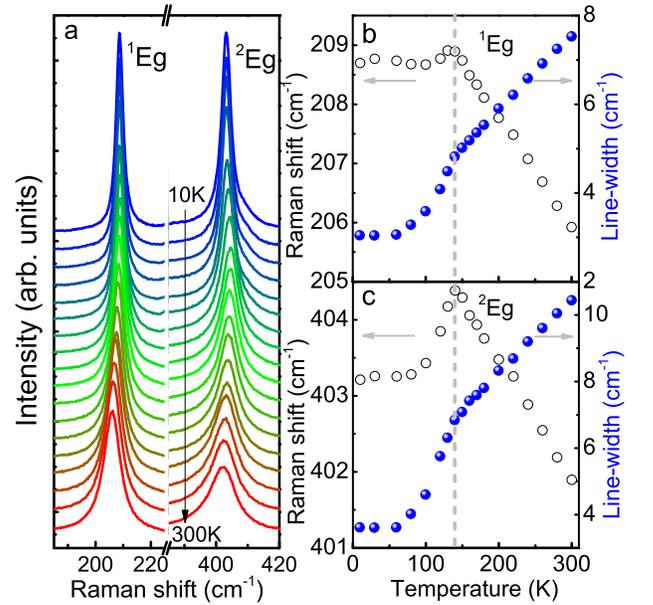}}
\caption{\label{fig3} (a) Temperature dependence of $^1E_{g}$ and $^2E_{g}$ phonon modes from 10 K to 300 K. (b)-(c) Temperature dependence of peak positions and line-widths of $^1E_{g}$ and $^2E_{g}$ modes.}
\end{figure}

\par
	We further carried out first-principles calculations. The observed modes can be well assigned to the calculated ones at zone center in both HT and LT phases (Table~\ref{table3}). The vibration patterns are classified and illustrated in Fig.~\ref{fig2} for the Raman-active modes. One of the four allowed $A_{1}$ mode given by calculations is absent in the spectra of LT phase. The missed $A_{1}$ mode with a calculated frequency of 212 cm$^{-1}$, is the vibration dominated by Li atom along c axis (Fig.~\ref{fig2}). The structural transition from $R\bar{3}c$ to $R3c$ in $\mathrm{LiOsO_3}$ is mainly attributed to the displacement of Li atom along the c axis\cite{Sim2014,Giovannetti2014}. The soft-mode theory requires that the soft modes related to the transition, if exist, should involve the vibrations of Li atom along the c axis. The above assignment of the phonon modes shows that the potential soft mode with Raman activity, exactly corresponds to $^1A_{1}$ mode. But unfortunately it is invisible in our Raman measurements. On the other hand, the absence of $^1A_{1}$ mode seems compatible with an order-disorder picture. The displacement magnitudes of Li atom in $\mathrm{LiOsO_3}$ and $\mathrm{LiNbO_3}$ are very close\cite{Veithen2002,Sim2014}. A strong soft mode at low temperature in $\mathrm{LiNbO_3}$\cite{Johnston1968} was observed while it is absent in $\mathrm{LiOsO_3}$. The clear contrast hints a different transition mechanism in $\mathrm{LiOsO_3}$ from displacive phase transition in $\mathrm{LiNbO_3}$. Recently, Liu et al. also argued\cite{Liu2015} that there may be no need for mode softening in $\mathrm{LiOsO_3}$ if it is an order-disorder transition and the double potential wells, between which Li atoms oscillate, are little changed crossing the transition.

\par
Temperature dependence of $^1E_{g}$ and $^2E_{g}$ phonon modes is shown in Fig.~\ref{fig3}a. And in Fig.~\ref{fig3}b and Fig.~\ref{fig3}c, we show the temperature dependence of peak positions and line-widths (full width at half maximum) of the two phonon modes, which have been fitted with Lorentzian functions. The anomalies in peak positions of both modes around 140 K, indicate that there occurs a structural transition. And the continuous changes of peak positions further suggests that the transition is a second-order type, consistent with the results of Ref. \onlinecite{Shi2013}. It should be noted that an unusually rapid and significant increase in line-widths below the transition temperature have been observed for the two modes. In fact, the similar behavior has also been observed in $\mathrm{SrSnO_3}$ \cite{Singh2008} and was attributed to a continuous order-disorder transition at 377 $^\circ$C.  For comparison, there is no such a significant increase in isostructural $\mathrm{SrHfO_3}$\cite{Singh2014}, which is believed to have a displacive transition at 1023 K. This implies that $\mathrm{SrSnO_3}$ and $\mathrm{LiOsO_3}$ may share the similar transition mechanism. Furthermore, detailed X-ray and neutron diffractions\cite{Shi2013} demonstrated that Li atoms are ordered in the LT phase, but disordered in the HT phase.  The unusual increase in line-widths is most likely caused by the order-disorder transition of Li atoms. The results allow one to deduce that the ferroelectric-like structural transition from $R\bar{3}c$ to $R3c$ in $\mathrm{LiOsO_3}$ is a continuous order-disorder transition.

\section{Summary}
In summary, we have carried out polarized and temperature-dependent Raman measurements on single crystal $\text{LiOsO}_\text{3}$. The spectra in the HT and LT phases are collected and a comparative study is made. The observed modes in both phases are assigned through a careful symmetry analysis and first-principles calculations. The anomalous increases in line-width and continuous changes of Raman frequencies around the transition temperature in several modes, seem to support the picture of a continuous order-disorder transition.

\begin{acknowledgments}
This work was supported by the Ministry of Science and Technology of China (973 project: 2012CB921701) and the NSF of China. Q.M.Z. and K.L. were supported by the Fundamental Research Funds for the Central Universities and the Research Funds of Renmin University of China. Y.G.S. was supported by the Strategic Priority Research Program (B) of the Chinese Academy of Sciences (Grant No. XDB07020100). Computational resources have been provided by the Physical Laboratory of High Performance Computing at Renmin University of China.

\end{acknowledgments}

\bibliography{Reference}

\end{document}